\documentclass{PoS}

\title{Two-flavor color superconductivity at finite temperature, chemical potential and in the presence of strong magnetic fields}

\ShortTitle{Magnetized two-flavor color superconductivity}

\author{\speaker{Sh. Fayazbakhsh}\\
        Institute for Research in Fundamental Sciences (IPM), School of Particles and Accelerators, P.O. Box 19395-5531, Tehran, Iran\\
        E-mail: \email{shfayazbakhsh@ipm.ir}}

\author{{N. Sadooghi}\\
        Department of Physics, Sharif University of Technology, P.O. Box 11155-9161, Tehran, Iran\\
        E-mail: \email{sadooghi@physics.sharif.ir}}

\abstract{Utilizing an extended two-flavor Nambu-Jona Lasinio (NJL) model, we review
 some of the effects of external magnetic fields on two-flavor color superconducting phase (2SC)
  at moderate baryon densities in the QCD phase diagram. The effective action of the extended NJL model consists
  of two mass gaps as functions of three intensive quantities, the temperature, the quark chemical
  potential and the external magnetic field. The nonzero values of the mass gaps,
  chiral and diquark condensates, induce spontaneous chiral and color symmetry breaking, respectively,
   and as a result two different phases of quark matter appear. We find the transition curves
    between these phases as well as the critical points in the QCD phase diagram in terms of
    the intensive quantities. Imposing a constant strong magnetic field on these two phases,
    we show that the mass gaps increase with the magnetic field and the symmetry breaking region
     in the QCD phase diagram expands even to the larger values of temperature and quark chemical
      potential. This phenomenon is a consequence of the magnetic catalysis of dynamical symmetry
       breaking, which is proven before.}

\FullConference{Xth Quark Confinement and the Hadron Spectrum,\\
        October 8-12, 2012\\
        TUM Campus Garching, Munich, Germany}

\begin{document}
\section{Magnetized Two-Flavor NJL Model at Finite T and $\mu$}
The extended fermionic Lagrangian density of a two-flavor gauged NJL
model is \cite{fayaz2011},
\begin{eqnarray}\label{eq1}
\lefteqn{\hspace{-7cm}{\cal{L}}_{f}=\overline{\psi}(x)[i\gamma^{\mu}(\partial_{\mu}-i\tilde{e}\tilde{Q}\tilde{A}_{\mu})+\mu\gamma^{0}]\psi(x)}\nonumber\\
\lefteqn{\hspace{-6.6cm}+G_{S}[(\overline{\psi}(x)\psi(x))^2+(\overline{\psi}(x)i\gamma_{5}\vec{\tau}\psi(x))^2]+G_{D}[(i\overline{\psi}^C(x)\epsilon_{f}\epsilon_{c}^{3}\gamma_{5}\psi(x))(i\overline{\psi}(x)
\varepsilon_{f}\epsilon_{c}^{3}\gamma_{5}\psi^C(x))],}
\end{eqnarray}
where, $\psi^C=C\overline{\psi}^T$ with $C=i\gamma^2\gamma^0$,
$\vec{\tau}=(\tau_{1},\tau_{2},\tau_{3})$ are Pauli matrices, and
$(\epsilon^{3}_c)^{ab}\equiv(\epsilon_c)^{ab3}$ and
$(\varepsilon_{f})_{ij}$ are antisymmetric matrices in color and
flavor spaces, respectively. Moreover, $G_S~\big(G_D\big)$ is the
scalar (diquark) coupling. The rotated electric charge operator in
the color-flavor space is, $\tilde{Q}=Q_{f}\otimes
1_{c}-1_{f}\otimes \left(\frac{\lambda_{8}}{2\sqrt{3}}\right)_{c}$
and the rotated photon field is
$\tilde{A}_{\mu}=A_{\mu}\cos\theta-G^8_{\mu}\sin\theta$. Here, $Q_f$
and $\lambda_8$ are the electric charge and the 8th Gell-Mann
matrices respectively, and, $A_{\mu}~\big(G^8_{\mu}\big)$ is the
photon (8th gluon) gauge field. Each quark degree of freedom has a
rotated electric charge in units of $\tilde{e}=e\cos\theta$ with
$\cos\theta=-\sqrt{3}g/\sqrt{3g^2+e^2}$, where e and g are the
electromagnetic and strong couplings, respectively. The rotated
charges are
$\left(\tilde{q}_{u_{r}}=1/2,\tilde{q}_{u_{g}}=1/2,\tilde{q}_{u_{b}}=1,\tilde{q}_{d_{r}}=-1/2,
\tilde{q}_{d_{g}}=-1/2,\tilde{q}_{d_{b}}=0\right).$ Introducing the
auxiliary fields, $\sigma=-2 G_{S}(\overline{\psi}\psi)$, as meson
and
$\Delta=-2G_{D}(i\overline{\psi}^C\varepsilon_{f}\epsilon_{c}^{3}\gamma_{5}\psi)$,
as diquark fields to the partition function corresponding to
(\ref{eq1}) and eventually integrating out the fermions, we obtain
the one-loop effective potential in terms of the mass gaps, $\sigma$
and $\Delta$, (see \cite{fayaz2011} for more details). Assuming that
the external magnetic field is aligned in the third direction,
$\tilde{\mathbf{B}}=B{\mathbf{e}}_{3}$, the magnetized effective
potential at finite $T$ and quark chemical potential $\mu$ is given
by
\begin{eqnarray}\label{eq2}
\lefteqn{\hspace{-7cm}\Omega_{\mbox{\tiny{eff}}}=\frac{\sigma^2}{4G_S}+\frac{|\Delta|^2}{4G_D}+\frac{B^2}{2}-\frac{2}{\beta}\int\frac{d^{3}p}{(2\pi)^{3}}\left\{\beta
E_{0}+\mbox{ln}\left(1+e^{-\beta(E_{0}+\mu)}\right)+\mbox{ln}\left(1+e^{-\beta(E_{0}-\mu)}\right)\right\}}\nonumber\\
\lefteqn{\hspace{-6.25cm}-\frac{2\tilde{e}B}{\beta}\sum^{+\infty}_{n=0}\alpha_{n}\int_{-\infty}^{+\infty}\frac{d
p_{3}}{8\pi^{2}}\left\{\beta
E_{+1}+\ln\left(1+e^{-\beta(E_{+1}+\mu)}\right)+\ln\left(1+e^{-\beta(E_{+1}-\mu)}\right)\right\}}\nonumber\\
\lefteqn{\hspace{-7cm}-\frac{4\tilde{e}B}{\beta}\sum^{+\infty}_{n=0}\alpha_{n}\int_{-\infty}^{+\infty}\frac{dp_{3}}{16\pi^{2}}\left\{\beta\left(|{E}^{+}_{\frac{+1}{2}}|+|{E}^{-}_{\frac{+1}{2}}|\right)+
2\ln\left(1+e^{-\beta{E}^{+}_{\frac{+1}{2}}}\right)+2\ln\left(1+e^{-\beta{E}^{-}_{\frac{+1}{2}}}\right)\right\}.}
\end{eqnarray}
Here,
$E_{\tilde{q}}=\sqrt{2|\tilde{q}\tilde{e}B|n+p_{3}^{2}+\sigma^{2}}$,
for $\tilde{q}=1,\pm\frac{1}{2}$, $E_{\tilde
{q}=0}=\sqrt{\vec{p}^{2}+\sigma^{2}}$,
$E_{\tilde{q}}^{\pm}=\sqrt{\left(E_{\tilde{q}}\pm\bar{\mu}\right)+|\Delta|^{2}}$
and $\beta\equiv T^{-1}$ are considered. We use the Ritus method
\cite{ritus}, to determine the fermion propagators, and introduce
the Landau levels, $n$, with degeneracies, $\alpha_n=2-\delta_{n0}$.
The vacuum expectation values of bosons are determined through
global minimization of the effective potential. For numerical
calculations, we use the parameters $\Lambda=653.3$ MeV,
$G_{S}=5.0163$ GeV$^{-2}$, and $G_{D}=\frac{3}{4}G_{S}$,
\cite{huang}.
\section{The Phase Diagram of Two-Flavor Quark Matter}
\begin{figure}[h]
\center
\includegraphics[width=4.75cm,height=3cm]{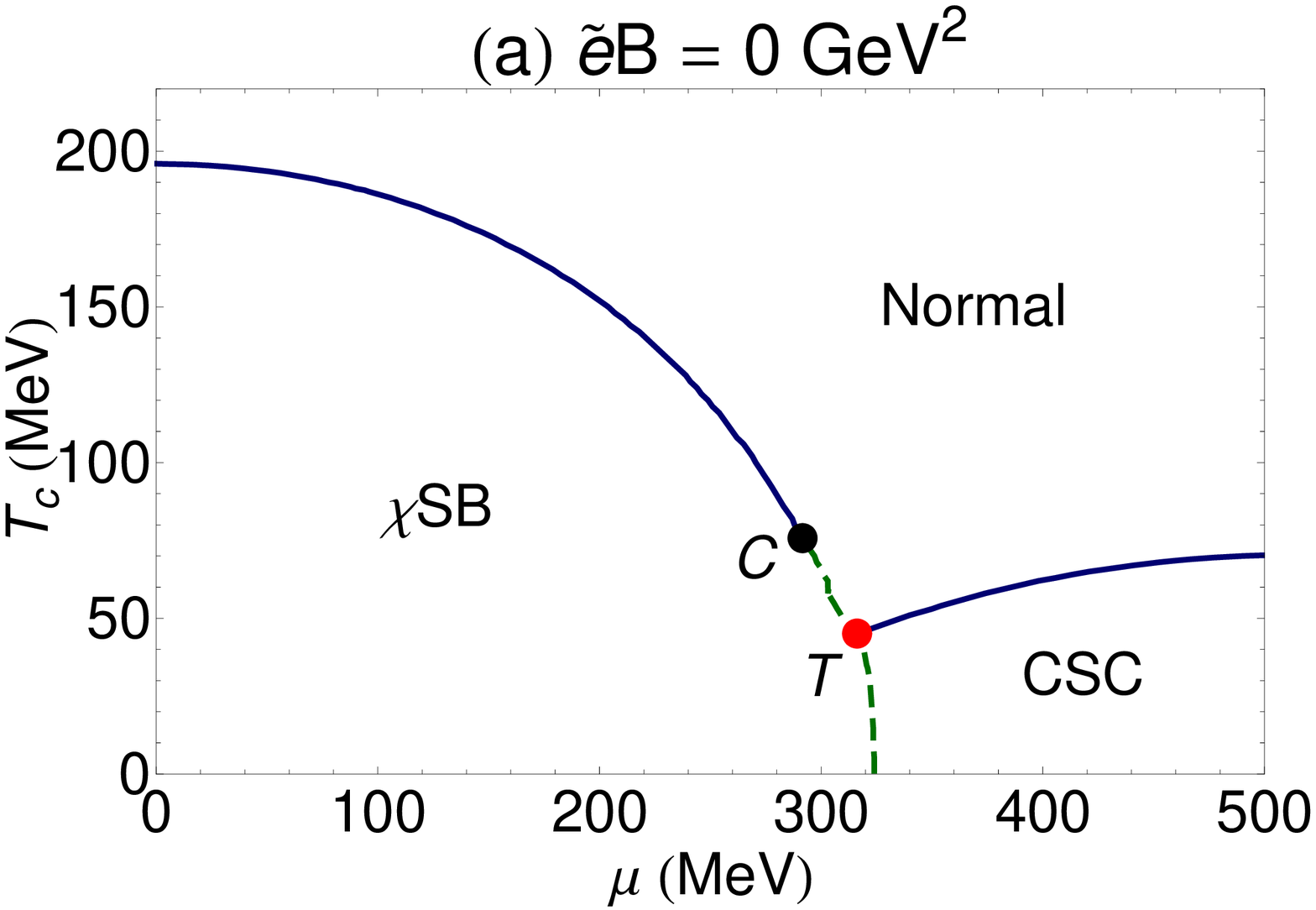}
\includegraphics[width=4.75cm,height=3cm]{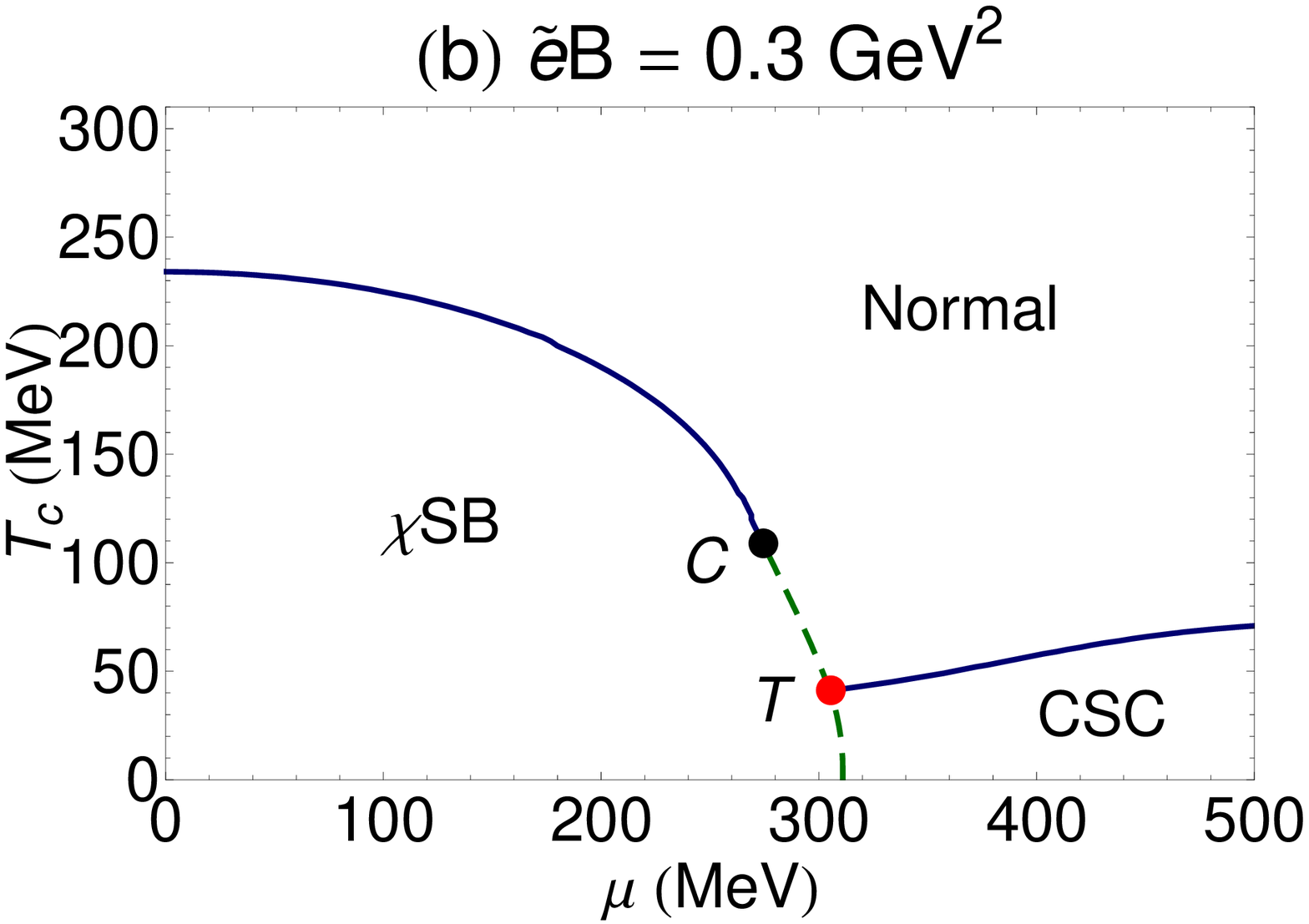}
\includegraphics[width=4.75cm,height=3cm]{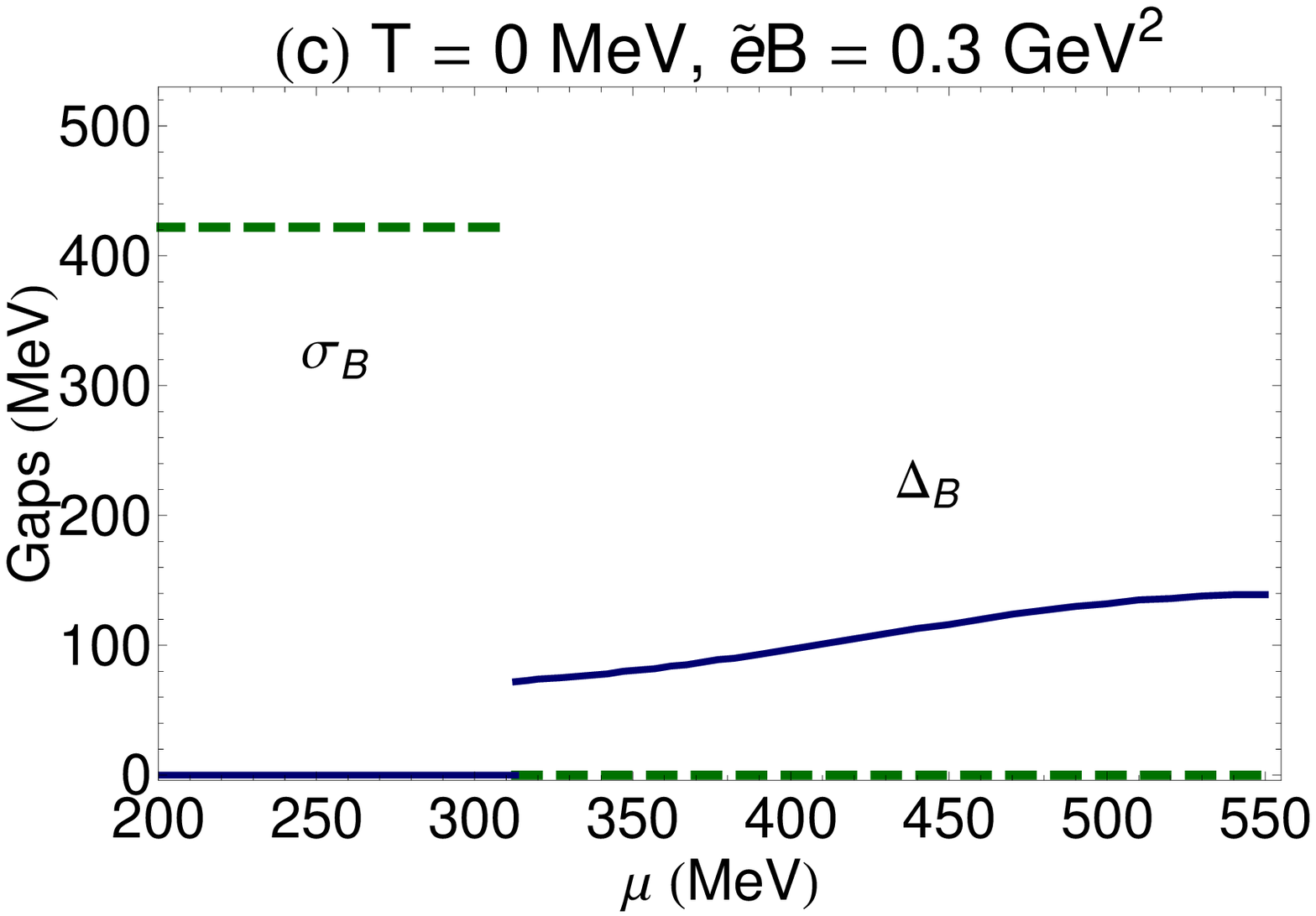}
\caption{$T_c-\mu$ phase diagram at (a) $\tilde{e}B=0$, (b)
$\tilde{e}B=0.3$ GeV$^2$. (c) Gaps in terms of $\mu$ at
$\tilde{e}B=0.3$ GeV$^2$.} \label{fig1}
\end{figure}

\begin{figure}[h]
\center
\includegraphics[width=4.75cm,height=3cm]{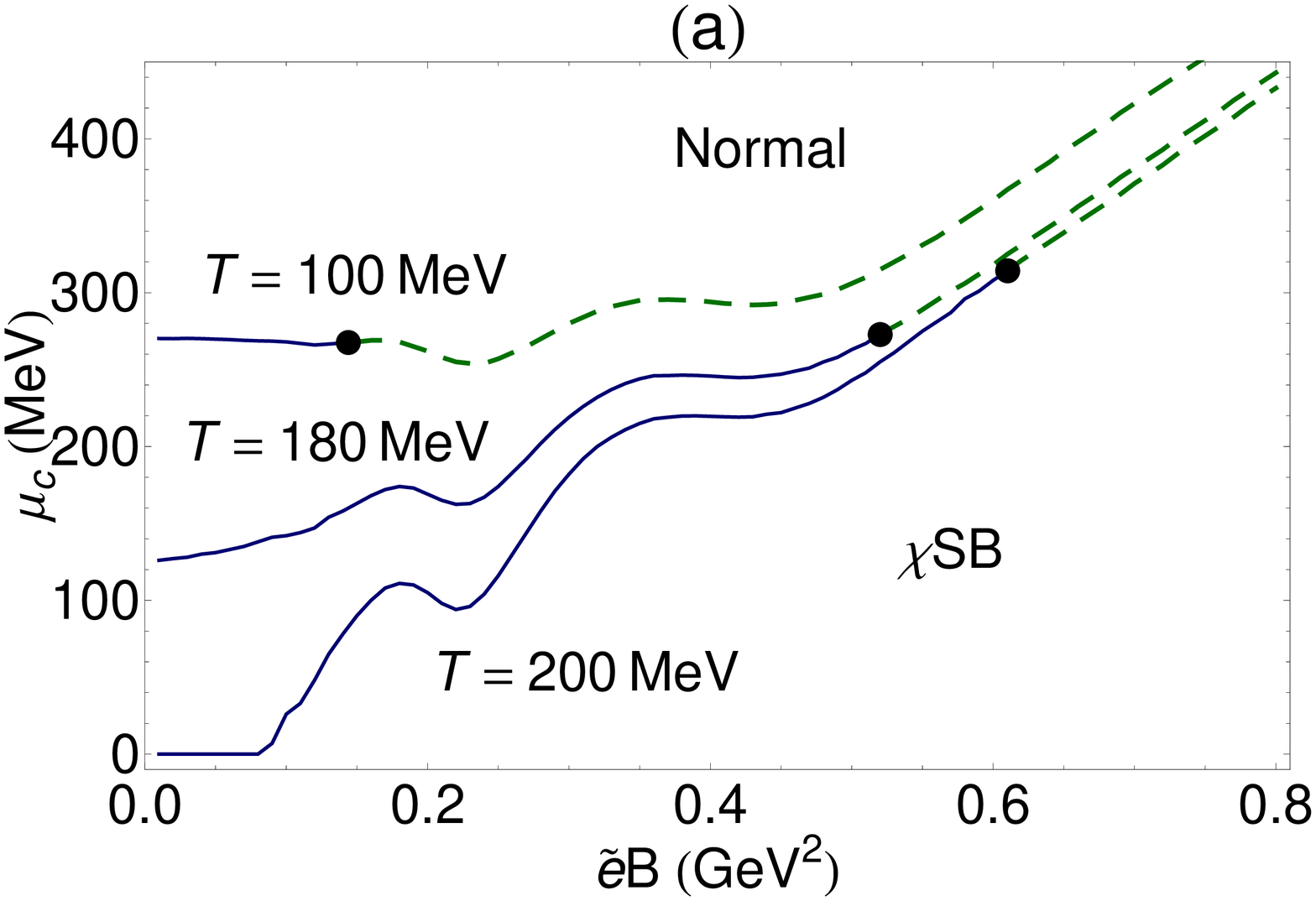}
\includegraphics[width=4.75cm,height=3cm]{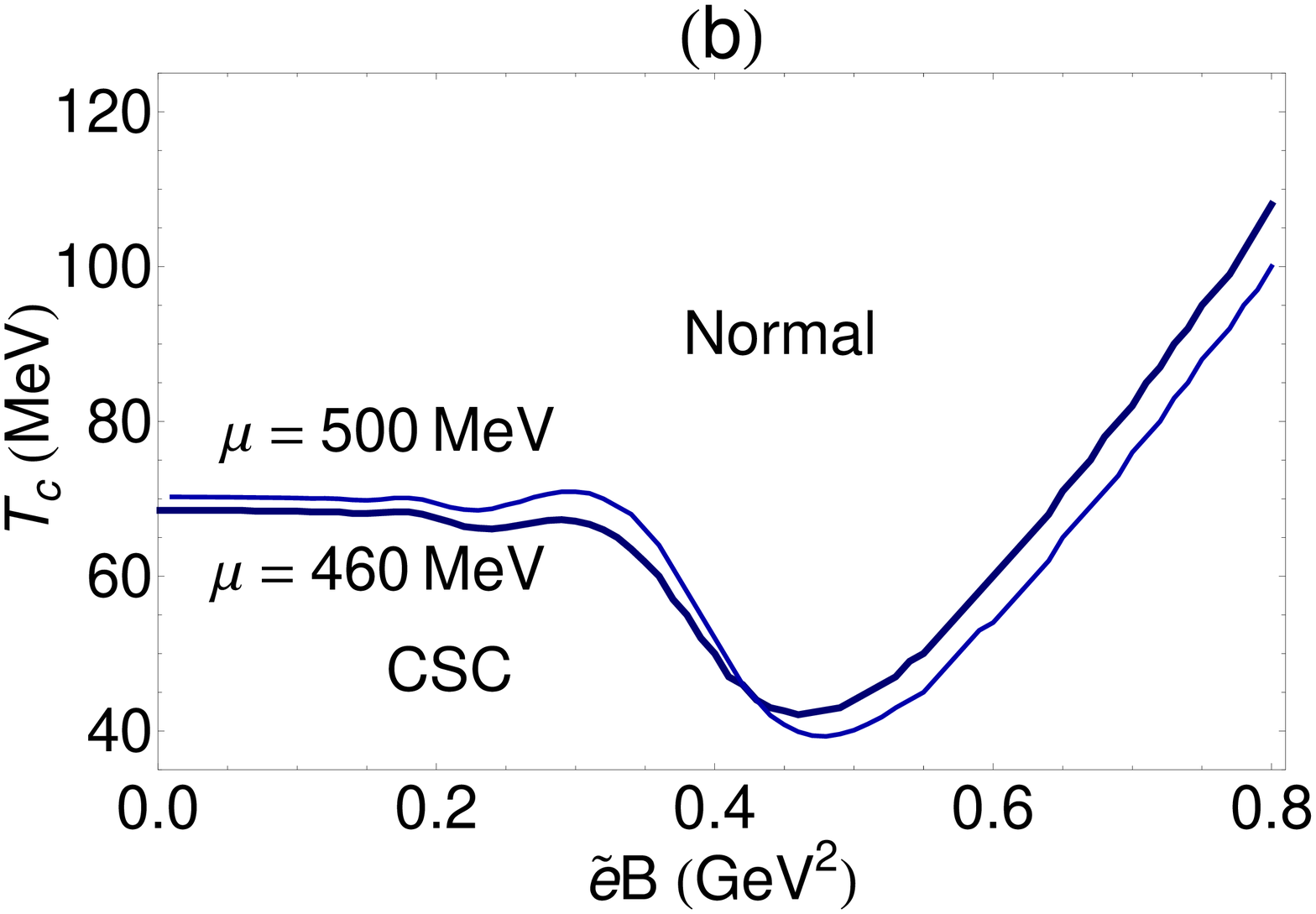}
\includegraphics[width=4.75cm,height=3cm]{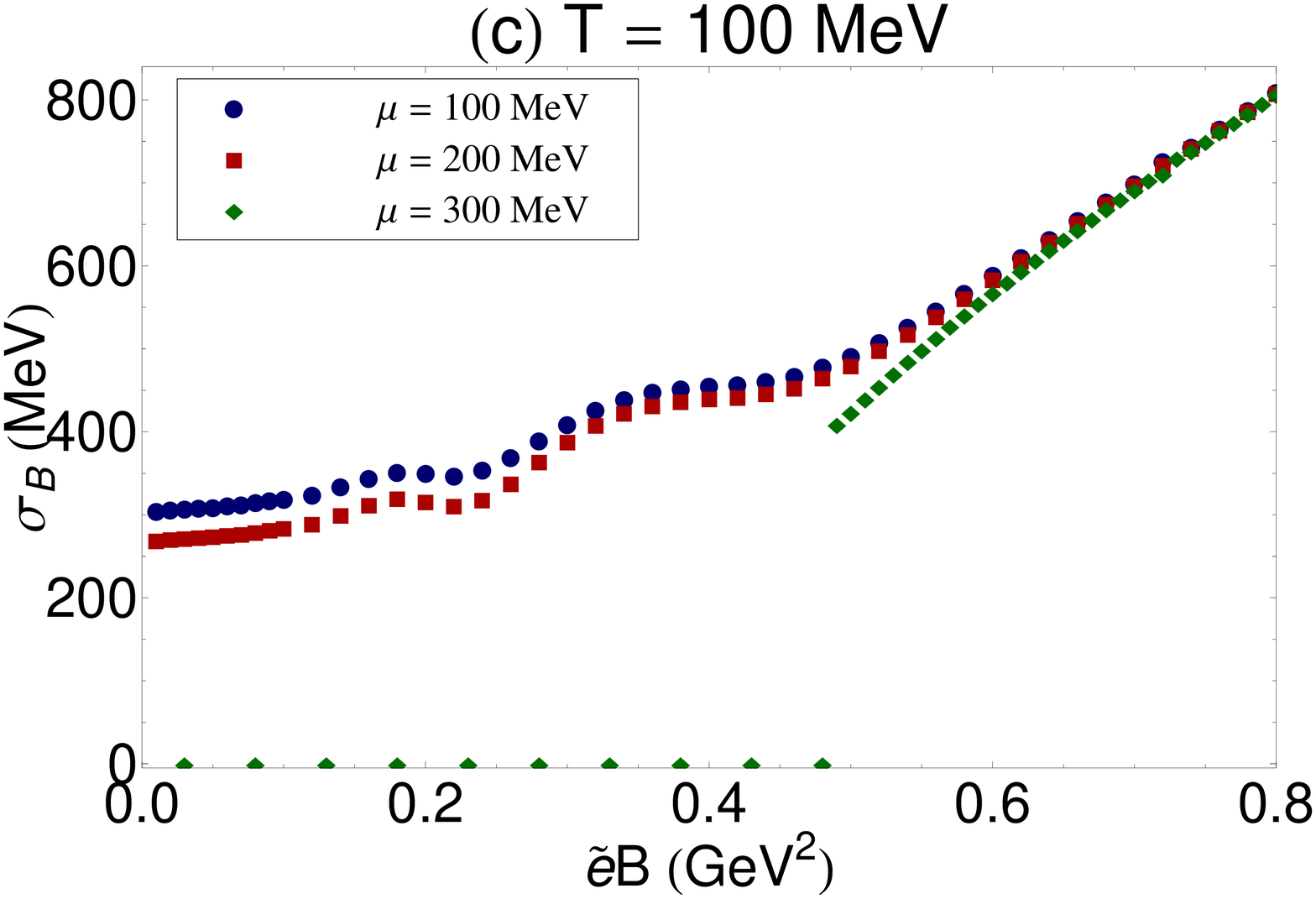}
\includegraphics[width=5.75cm,height=3cm]{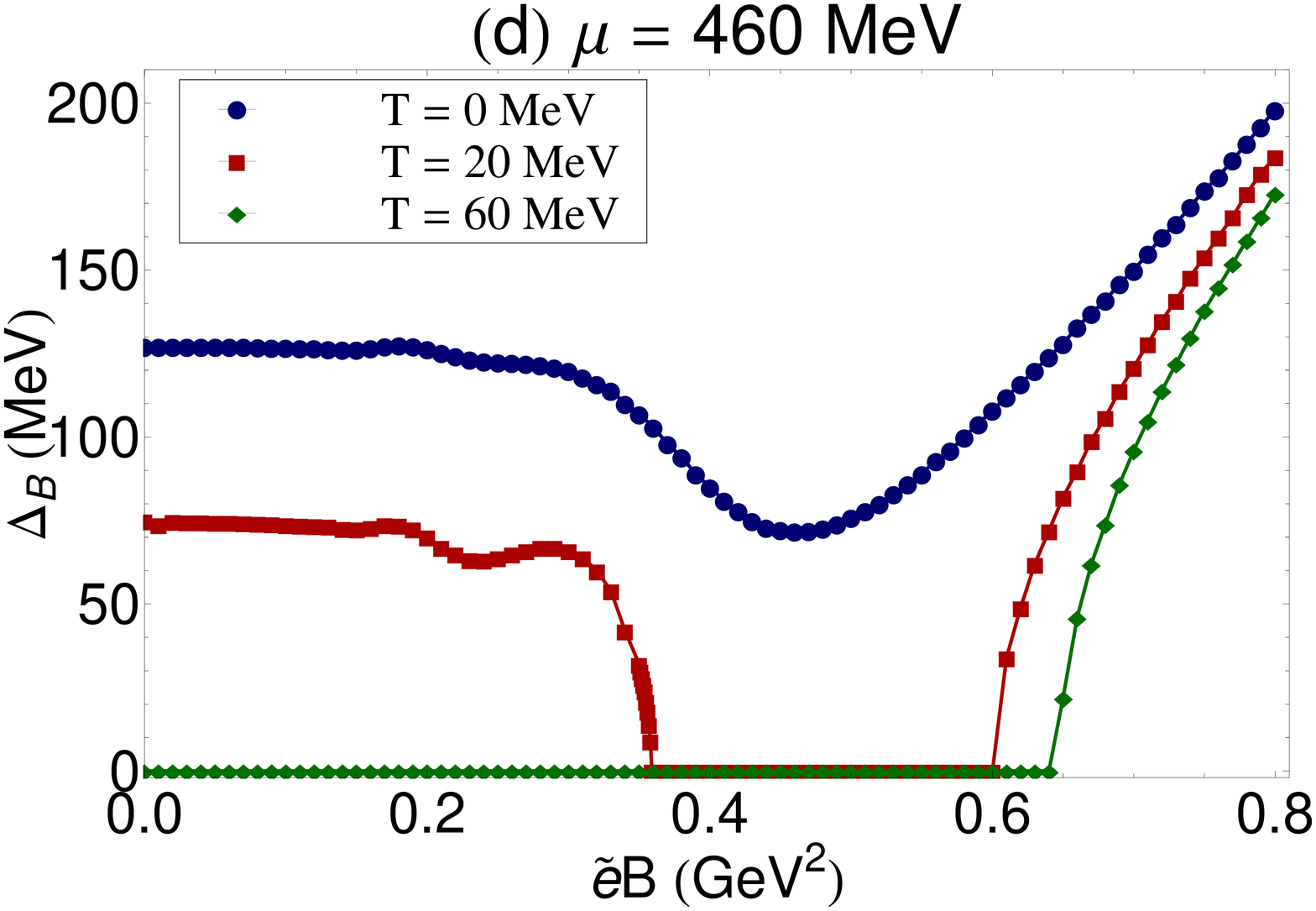}
\caption{$\tilde{e}B$ dependence of (a) $\mu_c$, (b) $T_c$, (c)
$\sigma_B$ and (d) $\Delta_B$ at some arbitrary constant values of
$T$ and $\mu$.} \label{fig2}
\end{figure}

The color-superconducting phase ($CSC$), where color symmetry is
spontaneously broken and the chiral symmetry broken phase ($\chi$SB)
are characterized for nonzero $\tilde{e}B$ by
$\sigma_B=0,\Delta_B\neq 0$, and $\sigma_B\neq 0,~\Delta_B=0$,
respectively. Defining these two phases and the normal phase by
$\sigma_B=0,~\Delta_B=0$, the two-flavor model phase diagrams in
three spaces, $T_c-\mu$, $\mu_c-\tilde{e}B$ and $T_c-\tilde{e}B$ are
shown in figures \ref{fig1}(a), (b) - \ref{fig2}(a), (b). The solid
(dashed) lines indicate the second (first) order phase transitions,
respectively. The dots C (T) denote the critical (tricritical)
points. In figure \ref{fig1}(c), the dependence of the mass gaps
$\sigma$ and $\Delta$ on $\mu$ at $(T,\tilde{e}B)=(0,0.3)$ GeV$^2$
confirms the first order transition between two phases shown in
\ref{fig1}(b).
According to the mechanism of magnetic catalysis, both $T_c$ and
$\mu_c$ grow with increasing $\tilde{e}B\gtrsim 0.45$ GeV$^2$. In
weak magnetic fields regime, $\mu_c$ and $T_c$ decrease with
increasing $\tilde{e}B$ in some regions of magnetic fields. This is
the phenomenon of ``Inverse Magnetic Catalysis'' discussed in
\cite{Inverse}. Van Alphen-de Haas oscillations are also
demonstrated in this regions. In Figs. \ref{fig2}(c) and
\ref{fig2}(d) the mass gaps are illustrated in terms of $\tilde{e}B$
at some arbitrary $T,\mu$ values. In \ref{fig2}(c) at $T=100$ MeV,
$\mu=300$ MeV a sudden jump in $\sigma_B$ from the zero value to a
nonzero one indicates a first order phase transition in
\ref{fig2}(a). A similar comparison will be conducted between
\ref{fig2}(b) and \ref{fig2}(d) at fixed $\mu=460$ MeV. For more
discussions see \cite{fayaz2011}.

%%%%%%%%%%%%%%%%%%%%%%%%%%%%%%%%%%%

\end{document}